\documentclass{JHEP3}
\usepackage{graphicx}
\usepackage{amsmath}

\renewcommand{\d}{\mbox{\boldmath$ d$}}

\renewcommand{\O}{\mbox{\boldmath$ O$}}
\newcommand{\Ttt}{\ensuremath{T_{3/2}}}

\newcommand{\Bpm}{\ensuremath{\mathcal{B}^{+-}}}
\newcommand{\Bpo}{\ensuremath{\mathcal{B}^{+0}}}
\newcommand{\Boo}{\ensuremath{\mathcal{B}^{00}}}
\newcommand{\Cpm}{\ensuremath{C^{+-}}}
\newcommand{\Spm}{\ensuremath{S^{+-}}}
\newcommand{\Coo}{\ensuremath{C^{00}}}

\newcommand{\ppld}{\ensuremath{\phi_{PLD}}}
\newcommand{\pes}{\ensuremath{\phi_{ES}}}
\newcommand{\ponei}{\ensuremath{\phi_{1i}}}

\newcommand{\tauz}{\ensuremath{\tau_{B^0}}}
\newcommand{\taup}{\ensuremath{\tau_{B^+}}}

\newcommand{\al}{\ensuremath{\alpha}}
\newcommand{\etal}{\mbox{et al.}}
\newcommand{\eg}{\mbox{e.g.}}
\newcommand{\ie}{\mbox{i.e.}}

\title{A Parameterization Invariant Approach to the Statistical
  Estimation of the CKM Phase $\alpha$}  

\author{Robin D. Morris\\USRA-RIACS, 444 Castro St, Suite 320, Mountain View, CA 94306, USA\\ E-mail: \email{rdm@riacs.edu} }
\author{Johann Cohen-Tanugi\\ Stanford Linear Accelerator Center, 2575 Sand Hill Rd, Mailstop 0029, Menlo Park, CA 94025, USA\\ and\\ Laboratoire de Physique Th\'eorique et Astroparticules, CNRS/IN2P3 et Universit\'e Montpellier 2,
Place Eug\`ene Bataillon, F-34095 Montpellier Cedex, France
\\ E-mail: \email{cohen@slac.stanford.edu} }

\abstract{
In contrast to previous analyses, we demonstrate a Bayesian approach
to the estimation of the CKM phase \al\ that is invariant to parameterization.
We also show that in addition  to {\em computing} the marginal posterior in a Bayesian
manner, the distribution must also be {\em interpreted} from a subjective
Bayesian viewpoint.  Doing so gives a very natural interpretation to
the distribution.

We also comment on the effect of removing information about \Boo.
}


\begin{document}



\section{Introduction}

A number of papers have been published recently that form a lively
debate about the nature of inference in particle physics in general,
and in the extraction of the CKM phase \al\ from measured
branching ratios and asymmetries in particular 
(see \eg\ \cite{babar07} and references therein 
for theoretical motivations and recent experimental results). 

The first paper, Charles \etal, \cite{charles2006}, proposed several
different parameterizations of the CKM phase \al\ problem and showed,
in their formulation, that different parameterizations resulted in
different posterior marginal distributions for \al.  These different
distributions were held to be the result of using flat priors in the
different parameterizations.  The interpretation of $p(\al)$ in
Charles \etal\ also claimed that it did not correctly identify the 8
known mirror solutions to the CKM phase \al\ problem.  Charles \etal\
also provided a simple 2-dimensional problem which they claimed showed
similar features.

Charles \etal\ is a criticism of the approach taken by the UTfit
collaboration \cite{bona2005}, and Bona \etal\ replied in
\cite{bona2007}.  In this paper the emphasis is shifted from full
distributions over \al\ to 95\% probability regions, which are shown
to be very similar to the 95\% confidence intervals given in Charles
\etal.  Bona \etal\ also note that the identification of the 8 modes
in the 1-CL plot of Charles \etal\ is not robust to slight changes in
the values of the observables, and that, in practice there is plenty
of information regarding the hadronic amplitudes which can (and
should) be used to remove some of the degeneracy.

Charles \etal\ replied in \cite{charles2007}, criticizing the change of
emphasis from $p(\al)$ to 95\% probability intervals as being an
admission that the approach of Bona \etal\ has significant dependence
on the parametrization chosen.  They also repeated their criticism
that the Bayesian marginal posterior, $p(\al)$ does not show the
expected 8-fold ambiguity.

A paper by Botella and Nebot \cite{botella2007} took another approach,
noting that some parameterizations used in the analysis of the CKM
phase \al\ problem are inadequate if they go beyond the minimal Gronau
and London assumptions \cite{gronau1990}.  In particular, the
``modulus and argument'' (MA) and ``real and imaginary'' (RI)
parameterizations of Charles \etal\ were shown to not uniquely identify
\al\ in the parameterization, leading to the leaking of spurious
information into $p(\al)$.  Botella and Nebot identified which
parameterizations do not suffer from this problem.  They also,
however, concentrated on probability regions, though they came
tantalizingly close to giving the correct Bayesian interpretation of
$p(\al)$ in their appendices C and E.

In this paper we will show how to perform a Bayesian analysis of the
problem that results in the same $p(\al)$ for {\em any}
parameterization.  We also show how regarding $p(\al)$ as a Bayesian
subjective distribution, \ie\ one that describes our state of
knowledge, allows it to be correctly interpreted in a straightforward
manner -- it is not sufficient just to use Bayes Theorem to perform
computation, the result of that computation must also be interpreted
from the Bayesian perspective.

We begin by reconsidering the simple 2-dimensional problem with mirror
solutions of Charles \etal\, as it is illustrative of some of the main
points we wish to make.

\section{\label{sec:2d}Mirror Solutions in a Simple 2D Problem}

The problem, from section VIII of \cite{charles2006}, is presented as
     {\em ``a theory predicts the expressions of two observables X and
       Y as functions of the two parameters $\alpha$ and $\mu$''}:
\begin{eqnarray}
X & = & (\alpha+\mu)^2 \nonumber \\
Y & = & \mu^2\, , \label{eq:simple}
\end{eqnarray}
where {\em ``an experiment has measured the observables from a
  Gaussian sample of events''} with the results:
\begin{eqnarray}
X & = & 1.00\pm 0.07 \nonumber \\
Y & = & 1.10\pm 0.07\, . \label{eq:xy}
\end{eqnarray}
In terms of the assumed physics, only $\alpha$ is of interest.  

It is important even at this early stage of the analysis to be clear
regarding what is considered an \lq\lq observable'', what is considered a
\lq\lq parameter'', and what is meant by saying that an observable has a
distribution, or that a parameter has a distribution.  Observables are
expected to have values that vary with different experimental data
sets, and saying that an observable has a distribution quantifies the
uncertainty due to a particular data set.  Saying that a parameter has
a distribution is a Bayesian concept, indicating that there is
actually a true, fixed, value, and that the distribution represents
our state-of-knowledge regarding what that value might be. 

This distinction is often somewhat artificial, however.  Typically the
quantities labeled as observables are not actually observed directly,
instead they are themselves inferred from observed data.  Different
data sets will give different distributions over the observables and,
consequently in the Bayesian framework, different distributions over
the parameters.  In equation \ref{eq:xy}, for example, the means and
variances for $X$ and $Y$ are the summary results of a particular data
set.

The standard approach to computing a joint Bayesian posterior distribution
for \al\ and $\mu$ is to use equations \eqref{eq:simple} and \eqref{eq:xy} to
define a likelihood, and then to combine it with a prior, $p(\alpha,\mu)$, on \al,
$\mu$,  giving
\begin{equation}
p_{\alpha,\mu}(\alpha,\mu|d) \propto
\frac{1}{2\pi\sigma_X\sigma_Y}
\exp\left(-\frac{[(\alpha+\mu)^2-\bar{X}]^2}{2\sigma_X^2}
-\frac{[\mu^2-\bar{Y}]^2}{2\sigma_Y^2}\right)
p(\alpha,\mu)
\label{eq:post1}
\end{equation}
where $d$ denotes the experimental data  and  $\bar{X}$,
$\bar{Y}$, $\sigma_X$ and $\sigma_Y$ are derived by considering the
full expression for the likelihood over the individual measurements.
They are all functions of $d$ \footnote{This simple for of the
  likelihood is a result of the assumed Gaussian errors.  In general,
  it will not be expressible in terms of summary statistics.}
\footnote{Conditioning 
  explicitly on the data, $d_i, i=1\ldots N_d$ Charles \etal's
  ``Gaussian sample of events'', gives 
  \[
  p(x|\d) \propto p(x) \prod_{i=1}^{N_d}
  \frac{1}{\sqrt{2\pi}\sigma_e}\exp\left(\frac{-(x-d_i)^2}{2\sigma_e^2}\right).  
  \]
  It is well known that the product of two Gaussians has variance {\em
    less than} either of the two.  As a consequence $p(\d|x)$ becomes
  steadily more peaked as more data is collected ($N_d$ increases).  The
  prior $p(x)$ does not change.   Thus, contrary to what is claimed in
  Charles \etal, it is often simple to show that {\em ``the
    relative prior dependence of the posterior distribution is reduced as
    the statistical information from the measured data is
    increased''.}}.

This formulation is subject to the standard criticism that different
parameterizations require different priors -- if, for example, we were to
parameterize the problem by \al, $\mu'$ where $\mu'=\mu^2$, then
clearly flat priors on $\mu$ and $\mu'$ will result in different
posterior distributions \cite{prosper2000}.

The discussion of observables and parameters above motivates an
alternative 
Bayesian analysis, one that results in a
posterior distribution that is invariant to the parameterization
chosen.  
In this analysis we first use the observed data to obtain a posterior
distribution over $X$ and $Y$.  This requires a prior on the
observables, and yields 
\begin{equation}
p_{X,Y}(x,y|d) \propto
\frac{1}{2\pi\sigma_X\sigma_Y} 
\exp\left(-\frac{(x-\bar{X})^2}{2\sigma_X^2}
-\frac{-(y-\bar{Y})^2}{2\sigma_Y^2}\right)
p(x,y)\, .
\label{eq:post2}
\end{equation}
Placing priors in the space of observables is reasonable: it is here
    that the experimenter will typically have good prior knowledge --
    prior knowledge that determined the design of the experiment.

The physical parameters of interest, \al, $\mu$ are related to $X$,
$Y$ by the deterministic relationships in equation \eqref{eq:simple}.
The distribution $p_{\al,\mu}(\al,\mu|d)$ is thus computed by the change of
variables rule. When the posterior for \al, $\mu$ is computed in this
way, the general result in Appendix \ref{app:invariance} can be used to show
that the resulting posterior marginal distribution, $p_{\al}(\al|d)$ is
invariant with respect to the chosen parameterization of the other
variables (in this case, $\mu$).

Changing variables gives
\begin{equation}
  p_{\alpha,\mu}(\alpha,\mu|d) \propto p_{X,Y}( x(\alpha,\mu),y(\alpha,\mu)|d)
  \left|\frac{\partial(X,Y)}{\partial(\alpha,\mu)}\right|
\end{equation}
resulting in
\begin{equation}
p_{\alpha,\mu}(\alpha,\mu|d) \propto
\frac{1}{2\pi\sigma_X\sigma_Y}
\exp\left(-\frac{[(\alpha+\mu)^2-\bar{X}]^2}{2\sigma_X^2}
-\frac{[\mu^2-\bar{Y}]^2}{2\sigma_Y^2}\right)
|\mu(\alpha+\mu)|
\label{eq:pauj}
\end{equation}
on the assumption of a flat prior $p(x,y)$, and 
where the factor of 4 is removed because of the multiple solutions.
This is plotted in figure \ref{fig:pua}.

Comparing equation \eqref{eq:pauj} with equation \eqref{eq:post1} it is
clear that this transformation of variables formulation is equivalent
to using the prior
\[
p(\al,\mu) \propto |\mu(\al+\mu)|.
\]
In this problem it is straightforward to show that the Jeffrey's
prior \cite{kass1996}, given by $\sqrt{|I(\al,\mu)|}$ where $I()$ is the Fisher
Information matrix,  is also proportional to $|\mu(\al+\mu)|$.  The Jeffrey's prior is
the prior that is invariant to transformation of the variables.  Thus,
computing a posterior $p_{X,Y}(x,y|d)$ using a uniform prior on $X$
and $Y$ followed by a transformation of variables to give
$p_{\al,\mu}(\al,\mu|d)$ is equivalent o using a Jeffrey's prior on
$\al$, $\mu$.

\FIGURE[t]{
\includegraphics[width=0.47\textwidth]{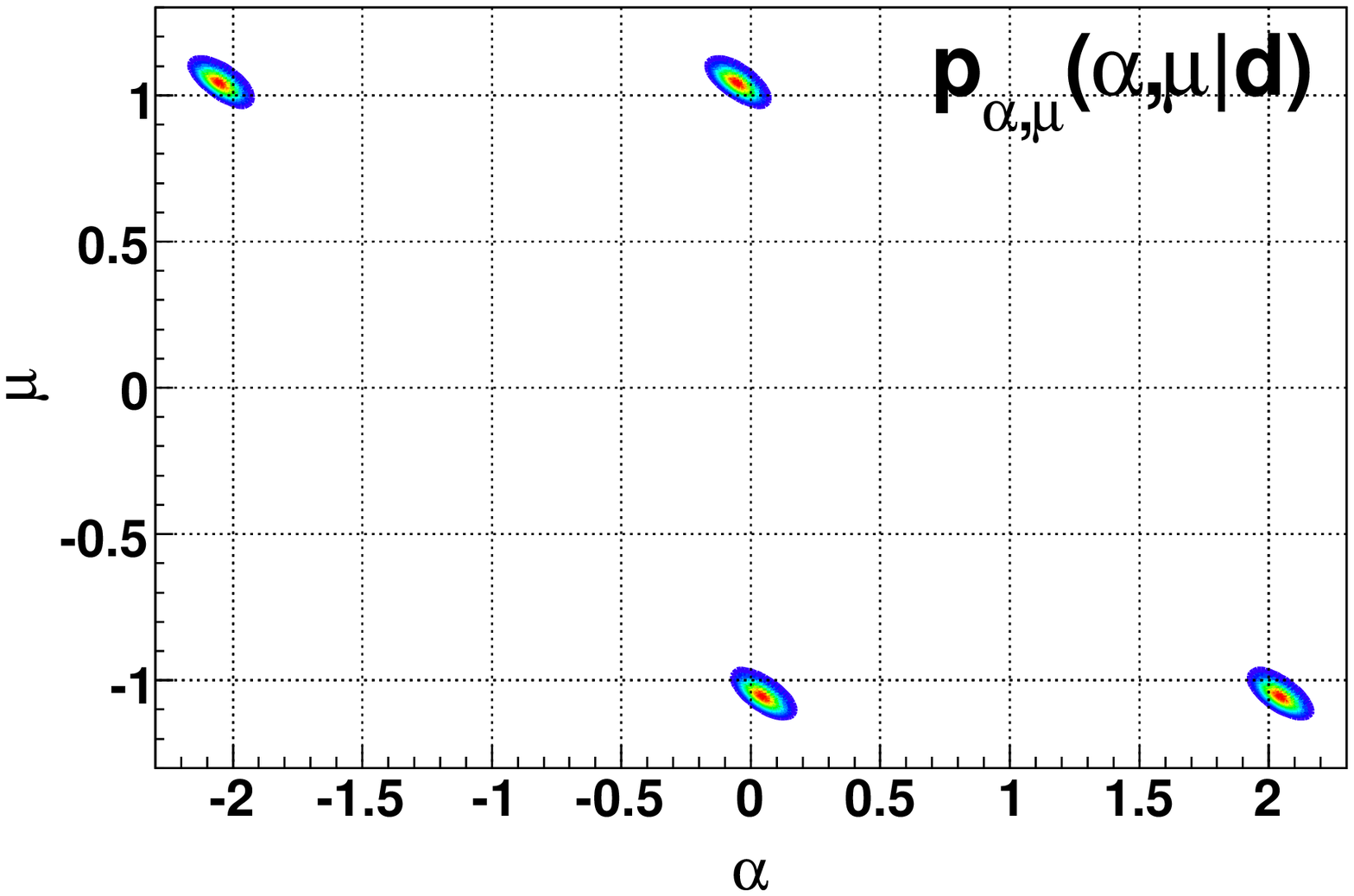}
\includegraphics[width=0.47\textwidth]{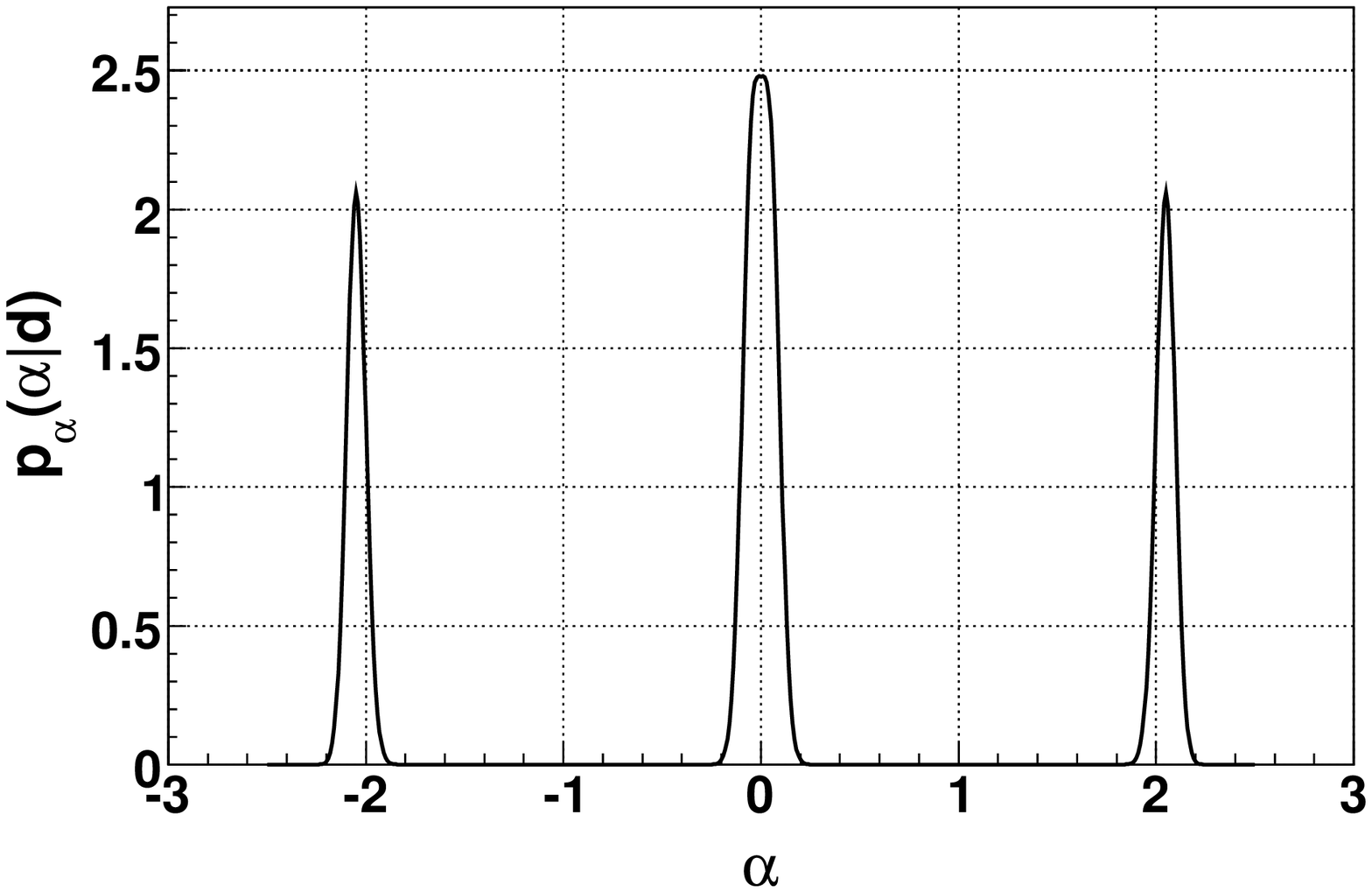}
\caption{\label{fig:pua}\label{fig:marginal}Left : The posterior
  distribution of $\mu,\alpha$; Right :  
  the marginal distribution $p_{\al}(\alpha|\d)$.}
}

While figure \ref{fig:pua} (left) looks very similar in projection to figure 5 in
\cite{charles2006}, note, however, that the modes of $p_{\alpha,\mu}(\alpha,\mu)$ are not
located at the values of $\alpha$ that were found by substituting the
mean values
$\bar{X}$ and $\bar{Y}$ into equations \eqref{eq:simple}.  They are
shifted because of the presence of the term $|\mu(\alpha+\mu)|$ in the
expression for $P_{\alpha,\mu}(\alpha,\mu)$ in equation
\eqref{eq:pauj}, coming from the determinant of the Jacobian of the
transformation from $X,Y$ to $\alpha,\mu$.  In this case the
displacement of the modes is small; it is not visible in figure
\ref{fig:pua}.  This need not be the case in general, and indeed is
not the case for the CKM phase \al\ problem.  See section
\ref{sec:ckm}.

The simplest way to form the marginal distribution $p(\al)$ is to
generate samples from the distributions of $X$ and $Y$, to
transform these samples into samples of $\alpha$ and $\mu$, and then
to plot a histogram of the samples of $\alpha$ \cite{smith1992}.
 In this case we generate
samples $x_i\leftarrow \mathcal{N}(1.00,0.07)$ and 
$y_i\leftarrow \mathcal{N}(1.10,0.07)$,
$i=1\ldots N$ for some suitably large $N$, and from each pair
$(x_i,y_i)$ we find the four solutions for $(\alpha_i,\mu_i)$, namely
\begin{eqnarray}
\alpha_i & = & \epsilon_x\sqrt{x_i} - \epsilon_y\sqrt{y_i} \nonumber
\\
\mu_i & = & \epsilon_y \sqrt{y_i}
\label{eq:sampleamu}
\end{eqnarray}
where $\epsilon_x=\pm 1$, $\epsilon_y=\pm 1$ and
each of the four $(\alpha_i,\mu_i)$ pairs is given weight
1/4\footnote{Note, however, that with finite 
  probability some of the samples $x_i$ and/or $y_i$ will be negative,
  resulting in imaginary values for $\mu_i$ and/or complex values for
  $\alpha_i$. This is not a problem with probability theory.  What it
  indicates is that the Gaussian distributions in equations
  \eqref{eq:xy} are only approximations to the true distributions of X
  and Y.}.

In the right panel of figure \ref{fig:marginal} we plot the marginal distribution
$p_{\alpha}(\alpha|\d)$, which is very similar to figure 6 (bottom) from
Charles \etal.  In their discussion of this figure, Charles
 \etal\ state that {\em ``if $\alpha$ and $\mu$ are fundamental
  physics parameters, Nature can only accommodate a single pair of
  values''}, and criticize the Bayesian approach by saying that the
 marginal $p_{\alpha}(\alpha|\d)$ only has 3 peaks, with the peak at zero being
 higher than the other two.  This is an incorrect interpretation of
 the distribution. This distribution is in fact
 exactly right when interpreted as a Bayesian subjective distribution,
 as representing {\em our state of knowledge}. 
 Nature has chosen one of the four modes visible in the joint
 distribution $p_{\al,\mu}(\al,\mu)$.  {\em We} do not know
 which one.  On the basis of {\em our} knowledge, 
 there are two chances out of four that Nature has
 chosen $\alpha\approx 0$, so our state of knowledge is {\em exactly}
 that $\alpha\approx 0$ is twice as likely as $\alpha\approx -2$ or
 $\alpha\approx 2$.  This is precisely what is shown by the
 distribution in the right panel of figure \ref{fig:marginal}, where the central mode has
 twice the area of each of the other two modes.

This simple problem has illustrated two of the key points we wish to
make, namely that the posterior distribution must be {\em interpreted}
in a subjective Bayesian manner, and that 
the posterior distribution in this type of problem
can be found by putting priors in the space of observables, and then
using the transformation of variables rule to compute the distribution
over the parameters derived from the observables.  The simple problem
is not rich enough to clearly demonstrate that this approach also leads to
posterior distributions for \al\ which are independent of the
parameterization chosen.  To do this, we turn now to the full CKM phase
\al\ problem.

\section{\label{sec:ckm}Extracting the CKM Phase $\alpha$}

There are six observable parameters involved in the CKM Phase
$\alpha$ problem, three CP averaged branching fractions, \Bpm, \Bpo,
\Boo, the direct CP asymmetries \Cpm and \Coo, and the $B^0\bar{B}^0$
mixing-induced CP asymmetry, \Spm.   These have been recently measured by
the B-factory experiments B{\sc a}B{\sc ar} and Belle
\cite{babar07, belle2006}.

The general formula for the branching ratio of a 2-body decay of a meson B can be found
in \cite{pdg} (eqs. 38.16 and 38.17). Specializing to a final state of light mesons, and 
averaging over CP-eigenstate yields:
\begin{eqnarray*}
  {\cal B}^{ij} & = & \frac{\tau_B^{i+j}}{16\pi
    M_B\hbar}\frac{|A^{ij}|^2+|\bar{A}^{ij}|^2}{2} \\
  C^{ij} & = & 
  \frac{|A^{ij}|^2-|\bar{A}^{ij}|^2}{|A^{ij}|^2+|\bar{A}^{ij}|^2} \\
  \Spm & = & \frac{2 Im (\bar{A}^{+-}A^{+-*})}{|A^{+-}|^2+|\bar{A}^{+-}|^2} \, .
\end{eqnarray*}
The decay amplitudes can be parameterized in a number of ways.  Here we
will consider three parameterizations, the Pivk-LeDiberder (PLD) and
Explicit Solution (ES) parameterizations considered in Charles \etal\ and
the so-called 1i parameterization from Botella and Nebot.  These vary
in how they parameterize $A^{ij}$ and $\bar{A}^{ij}$, but all include
\al\ explicitly as one of the parameters.  Details of the
parameterizations are given in appendix \ref{app:parameterizations}.

Denote the parameterizations as $(\al,\ppld)$, $(\al,\pes)$ and
$(\al,\ponei)$, where \ppld\ denotes the other five parameters of the PLD
parameterization, and similarly for \pes\ and \ponei.  Denote by \O\
the set of six observables, \Bpm, \Boo, \Bpo, \Cpm, \Coo\ and \Spm.
Then we have
\begin{eqnarray*}
  \O & = & f(\al, \ppld) \\
     & = & g(\al, \pes) \\
     & = & h(\al, \ponei)\ ,
\end{eqnarray*}
where the functional forms of $f()$, $g()$ and $h()$ 
can be derived from the parameterizations 
given in
Appendix \ref{app:parameterizations}.  Table \ref{tab:observables} gives the 
 values for the observables and their uncertainty that are used in this work \footnote{The values for the observables
  given in Table \ref{tab:observables} are those used in \cite{charles2006}, 
  as we wish to compare our method with theirs.  Subsequent
  improved measurements result in the distributions only having four
  modes.  See Appendix B of \cite{botella2007}.}.
Using a
uniform prior in the space of observables, these define a multivariate
Gaussian posterior, $p(\O|d)$ where $d$ is the experimental data.

\TABLE[t]{
  \begin{tabular}{|l|c|c|c|}
    \hline
    Observable    & \Bpm  & \Bpo & \Boo \\
    \hline
    Mean$\pm$std  & $(5.1\pm 0.4)\times 10^{-6}$ &
    $(5.5\pm 0.6)\times 10^{-6}$ &
    $(1.45\pm 0.29)\times 10^{-6}$ \\
    \hline
    Observable & \Cpm & \Coo & \Spm \\
    \hline
    Mean$\pm$std  &
    $-0.37\pm 0.10$ &
    $-0.28\pm 0.40$ &
    $-0.50\pm 0.12$ \\
    \hline
  \end{tabular}
  \caption{\label{tab:observables}World average values for the observables,  from
  \cite{charles2006}.}
}

Using the change-of-variables formulation gives
\[
p_{PLD}(\al,\ppld) = p( f(\al,\ppld) |d )|J_f|
\]
and the marginal distribution for \al\ is given by
\begin{equation}
  p_{PLD}(\al) = \int_{\ppld} p( f(\al,\ppld) |d )|J_f| d\ppld.
  \label{eq:ppld}
\end{equation}
Similarly
\begin{equation}
  p_{1i}(\al) = \int_{\ponei} p( h(\al,\ponei) |d )|J_h| d\ponei.
  \label{eq:p1i}
\end{equation}

In appendix \ref{app:invariance} we show that under reasonable
conditions these marginal distributions are identical, \ie\ that the
marginal posterior distribution for \al\ is {\em independent} of the
chosen parameterization.  This should not be surprising -- the same
information on the same observables gives the same information about
the same physical parameter.

\FIGURE[t]{
  \includegraphics[width=0.45\textwidth]{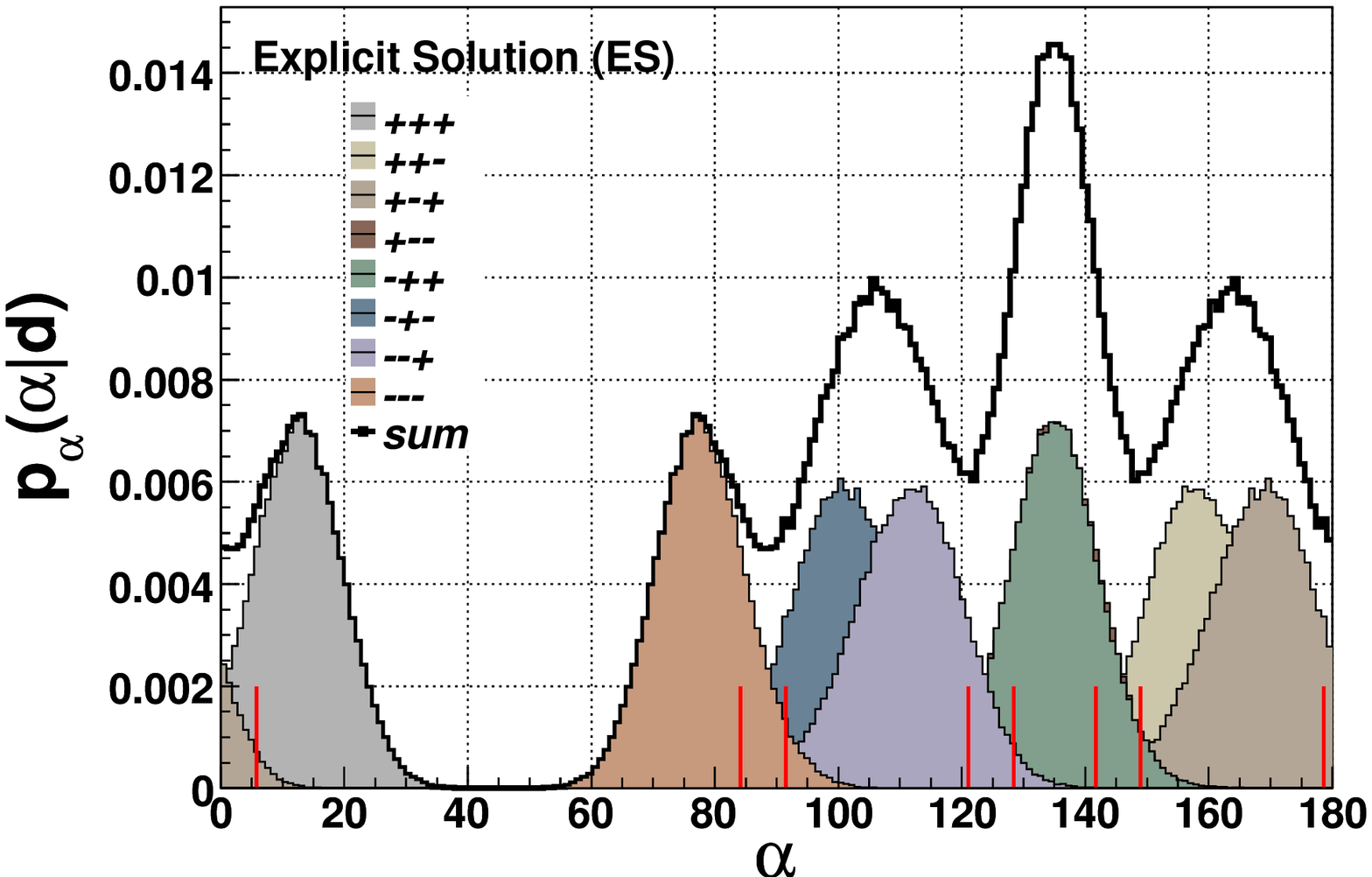}
  \includegraphics[width=0.45\textwidth]{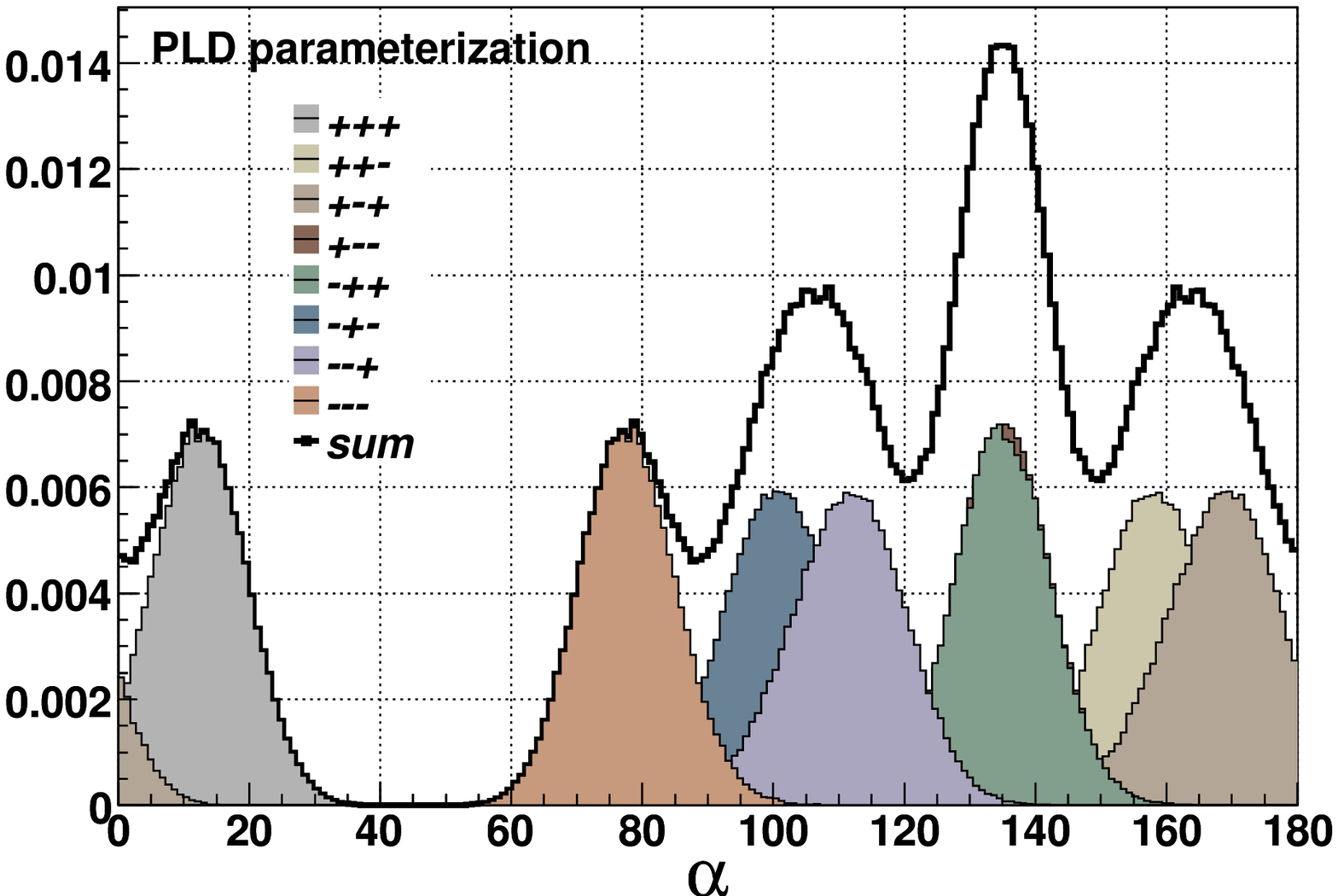}
  \includegraphics[width=0.45\textwidth]{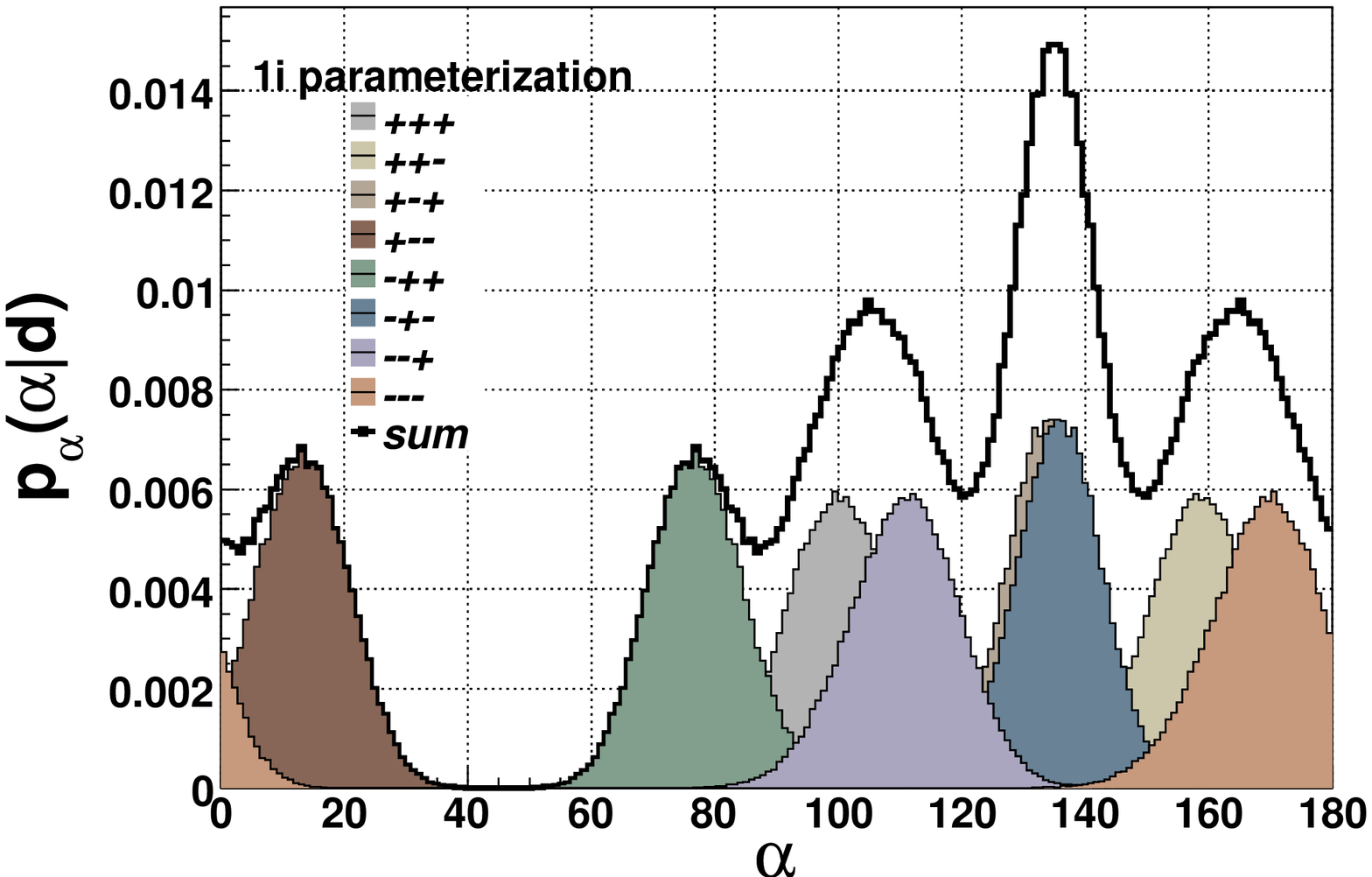}
  \includegraphics[width=0.45\textwidth]{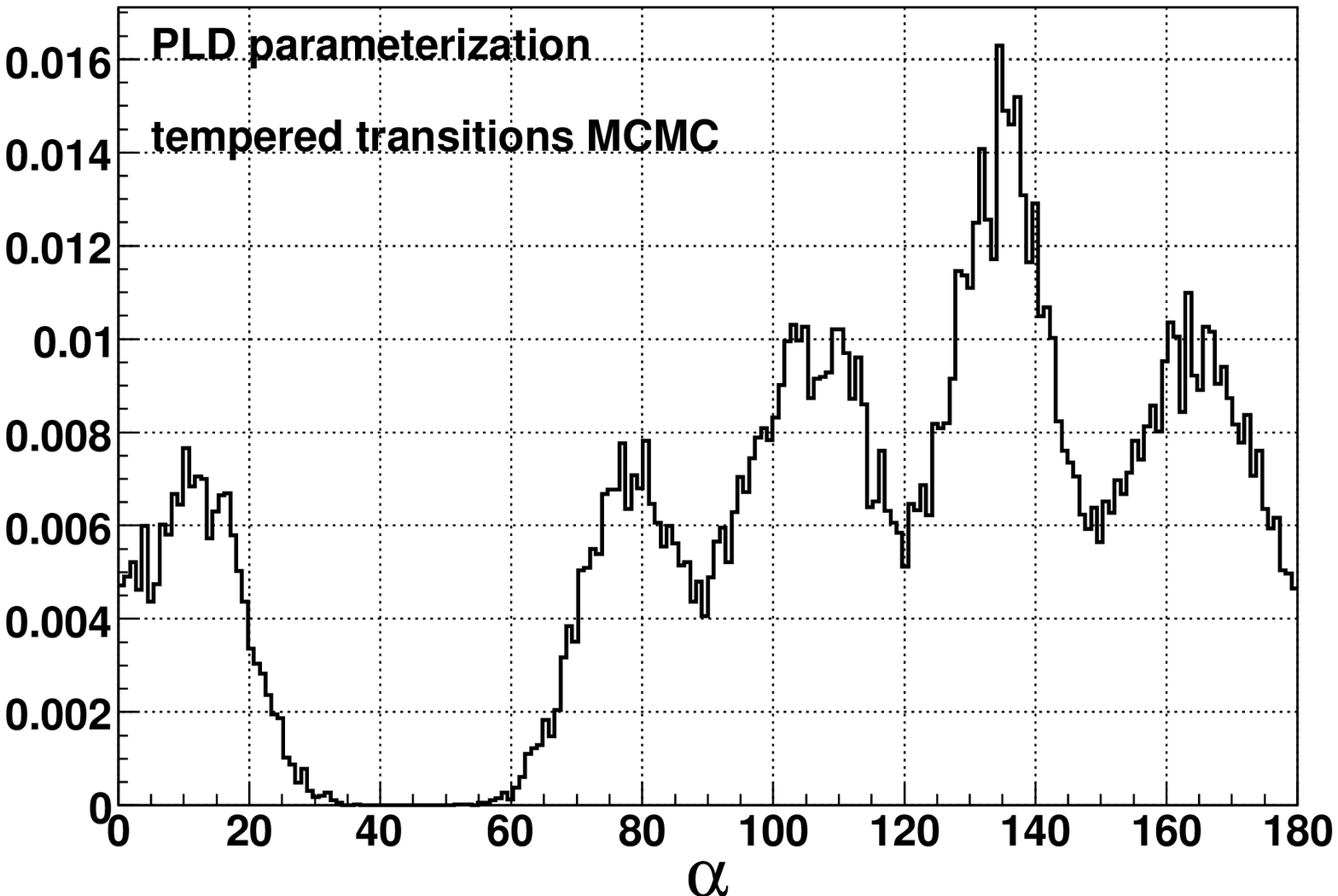}
  \caption{\label{fig:posterior}The first three plots show the marginal posterior
    distributions for \al\ under the PLD, 
    ES and 1i parameterizations generated by inverting the systems. 
The short vertical red lines on the top left plot indicate the central values obtained by \cite{charles2006}.
Legends indicate the tuple of signs corresponding to each mode.
    The final plot is of samples generated from the PLD
    parameterization using the tempered transitions MCMC
    scheme. Binning in $\alpha$ is identical for all figures.  
    In the first three plots, a sample of 100000 sets of observables is drawn, 
    and the choices of signs, as indicated by the legends, 
   allows each mode to be determined separately. As a result, the sum histograms 
    have 800000 non independent entries. The fourth histogram is of size 100000.}
}

In figure \ref{fig:posterior} we plot histograms
representing the three marginal posterior distributions.  The samples
were generated by sampling the observables and inverting the systems \footnote{If we choose
  to use non-flat priors on the observables., then we 
  can generate samples representing the distribution 
  $p(\alpha,\phi)$ by generating samples
  from the observables, weighting each sample by the prior, and then
  re-sampling the set of weighted samples to give samples from the
  posterior. See \cite{smith1992} for details.}.
As expected, the three histograms
are essentially identical.  We also show a histogram of samples generated using
the PLD parameterization and a Markov chain Monte Carlo algorithm
\cite{robert}.  As expected, the histogram is the same as the
others.  It is included to demonstrate that our approach is not
restricted to cases where the system can be inverted.  Care must be
taken in choosing the MCMC scheme, as the distribution is strongly
multimodal.  We used the simulated tempering scheme of \cite{neal1996}
which successfully sampled the 8 modes of the distribution.

If we consider the modulus-and-argument (MA) parameterization
\cite[equation 4]{bona2005}, we will not, however recover the same
distribution for $\alpha$.  The determinant of the Jacobian for the MA system is
identically zero for $\alpha=0$, and this results in a spurious zero
in probability at $\alpha=0$, the remainder of the distribution being
identical to our figure \ref{fig:posterior}.  This adds to the
discussion in \cite{botella2007} that the MA parameterization, by
going beyond the minimal Gronau and London assumptions, is unsuited to
the analysis of this problem.

The histograms generated by inverting the systems are 
clearly composed of 8 modes, one for each of the 8
solutions. (There are two modes that overlap almost totally around
$\alpha\approx 140^\circ$.)   By construction, each of these modes has equal probability
mass (=1/8), even though they are different shapes; the heights and
widths vary, but the area beneath each mode is the same.  Each
possible solution for $\alpha$ has different uncertainty (due to the
complex relationship between $\alpha$ and the observables), but each mode has
equal probability to be the one chosen by Nature\footnote{The reader
  is reminded that we are reconsidering the case discussed in
  Charles \etal.  A complete analysis of the CKM phase $\alpha$
  problem would include additional information which would break the
  symmetry \cite{charles05}.}\footnote{In this case, and in the 2d problem
  in section \ref{sec:2d}, it is known by construction that each mode
  contains the same proportion of the total probability (1/4 for each
  mode in the 2d problem and 1/8 for the CKM phase \al\ problem).  In
  general, however, this may not be known in advance.  Using a
  numerical search routine with random restarts can be used to locate
  the modes, and the Hessian, $H$, at each mode can be computed.  (Often
  this will be computed as a by-product of the numerical
  optimization.)  The probability volume in each mode can be
  approximated by $p(\hat{\theta})/\sqrt{\det(H/2\pi)}$ where
  $\hat{\theta}$ are the parameters at the mode \cite{sivia}.
  Alternatively, samples generated without knowing how many modes are
  present (\eg\ by using the tempered transitions MCMC scheme) can be
  clustered, and the number of samples in each cluster gives a measure
  of the probability volume in that mode.}.

The final marginal distribution is the sum of these 8 modes, which is
plotted as the dotted line.  This shows a large peak around
$\alpha=140^\circ$ and a number of smaller peaks.  Again, this distribution
correctly describes our state of knowledge -- there are 2 of the 8
modes near $\alpha=140^\circ$ and, because we don't know which mode Nature
has chosen, there are thus 2 chances out of 8 that $\alpha\approx
140^\circ$.  There is only 1 chance out of 8 that $\alpha\approx 80^\circ$, so the
peak there has half the area of the peak at $\alpha=140^\circ$.  This 
accurately represents our state of knowledge about \al.

Also shown on figure \ref{fig:posterior} are short vertical lines marking
the values of $\alpha$ that are found when the mean values for
the observables are transformed into the different parameterizations.
Again, it comes as no
great surprise that the mean of the distribution of the inputs is not
transformed to the mean of the distribution of the output, especially
when the uncertainty on some of the variables is of the same order as
the value itself, and the system of equations is highly
nonlinear\footnote{We note, however, that as the variances of the observables
  are reduced, the mean values remaining fixed, that the modes
  do converge to the values given by inverting the mean values.}.
This also naturally explains why there is still finite probability
density that $\alpha=0/180^\circ$. 

As the methodology presented in this work relies on the one-to-one
relationship (up to discrete ambiguities) between the observables
\{\Bpm, \Bpo, \Boo, \Cpm, \Coo, \Spm\} and the underlying isospin
amplitude representation, the analysis of the case when \Boo and \Coo
are not measured is not in general possible, once the system has been
inverted. For instance, although the PLD representation presents the
very appealing feature that $\alpha$ appears in the system
\eqref{eq:pld} only in the expressions for \Boo and \Coo, and therefore
cannot be determined when the latter are not measured, this feature is
not obvious anymore in the inverted system \eqref{eq:pldinv}.  This is
equivalent to the fact, already mentioned in Botella and Nebot
(section C.1), that \{\Boo, \Coo\} are algebraically constrained by
any set of measurements \{\Bpm, \Bpo, \Cpm, \Spm\} {\it and} the
assumption of isospin symmetry. As noted by Botella and Nebot, sampling
\Coo\ uniformly  between $-1$ and $+1$, and \Boo\ between $0$ and
$B_{max}$, results in a distribution that is much flatter than those
shown in figure \ref{fig:posterior}.  This distribution does not,
however, become flat as $B_{max}\rightarrow\infty$, 
because ultimately the shape of the underlying
single mode distributions will be driven by the algebraic constraints from the isospin assumption and by
the error propagation from the {\it measured} observables. As an illustration, we show in figure
\ref{fig:nopi0pi0} the result of the \lq 1i' parameterization for
$B_{max}=20\Boo$. Increasing the upper bound on \Boo\ will not change the final 
distribution, but will result in more samples being thrown away as incompatible with the constraints on the system. 
Figure \ref{fig:nopi0pi0} (right) shows a histogram of the samples of \Boo\ and \Coo\
that were retained.  It shows the probabilistic constraints on \Boo\ and \Coo\ due to the 
observations and the assumption of isospin symmetry.

\FIGURE[t]{
  \includegraphics[width=0.45\textwidth]{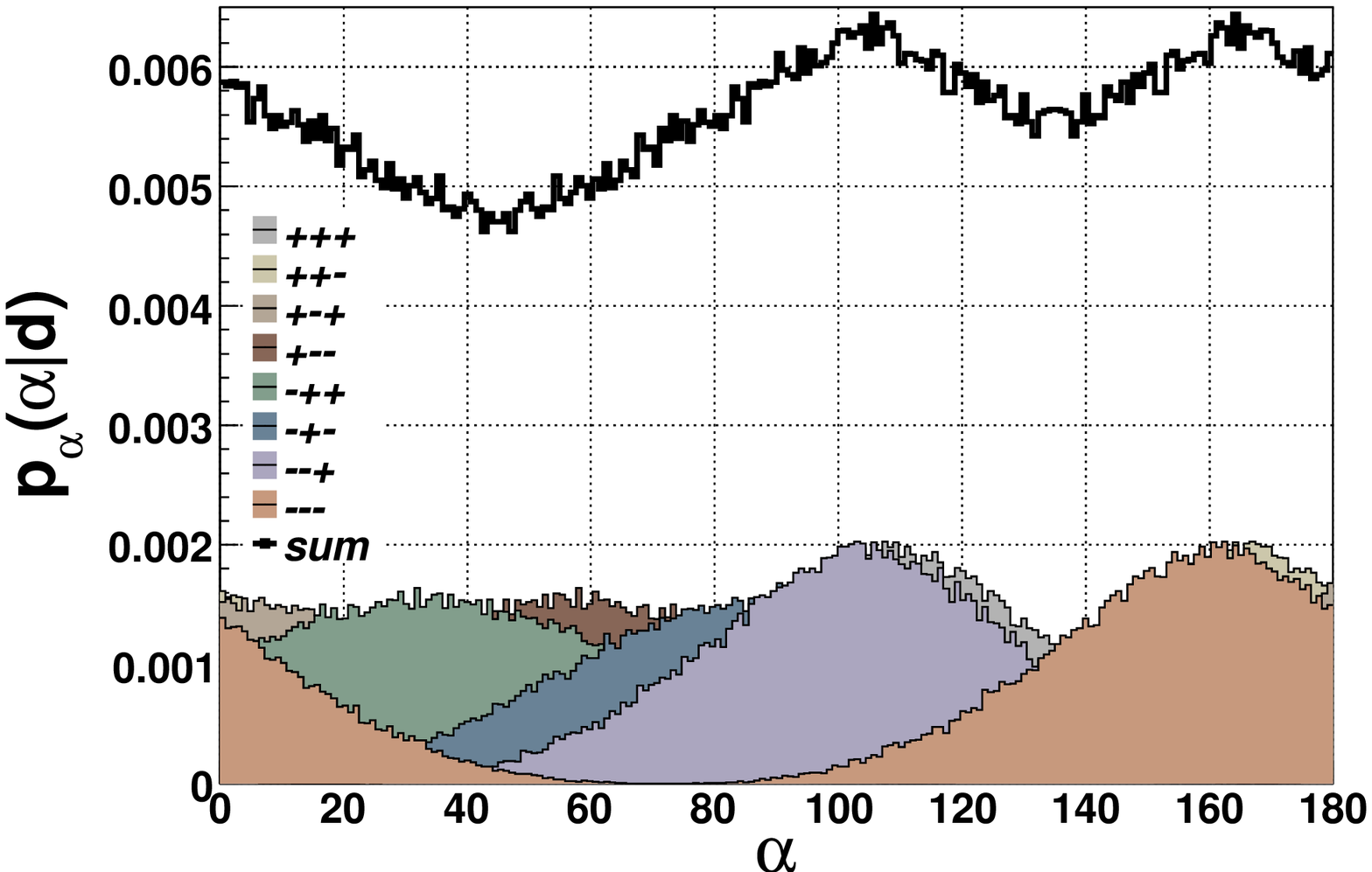}
  \includegraphics[width=0.45\textwidth]{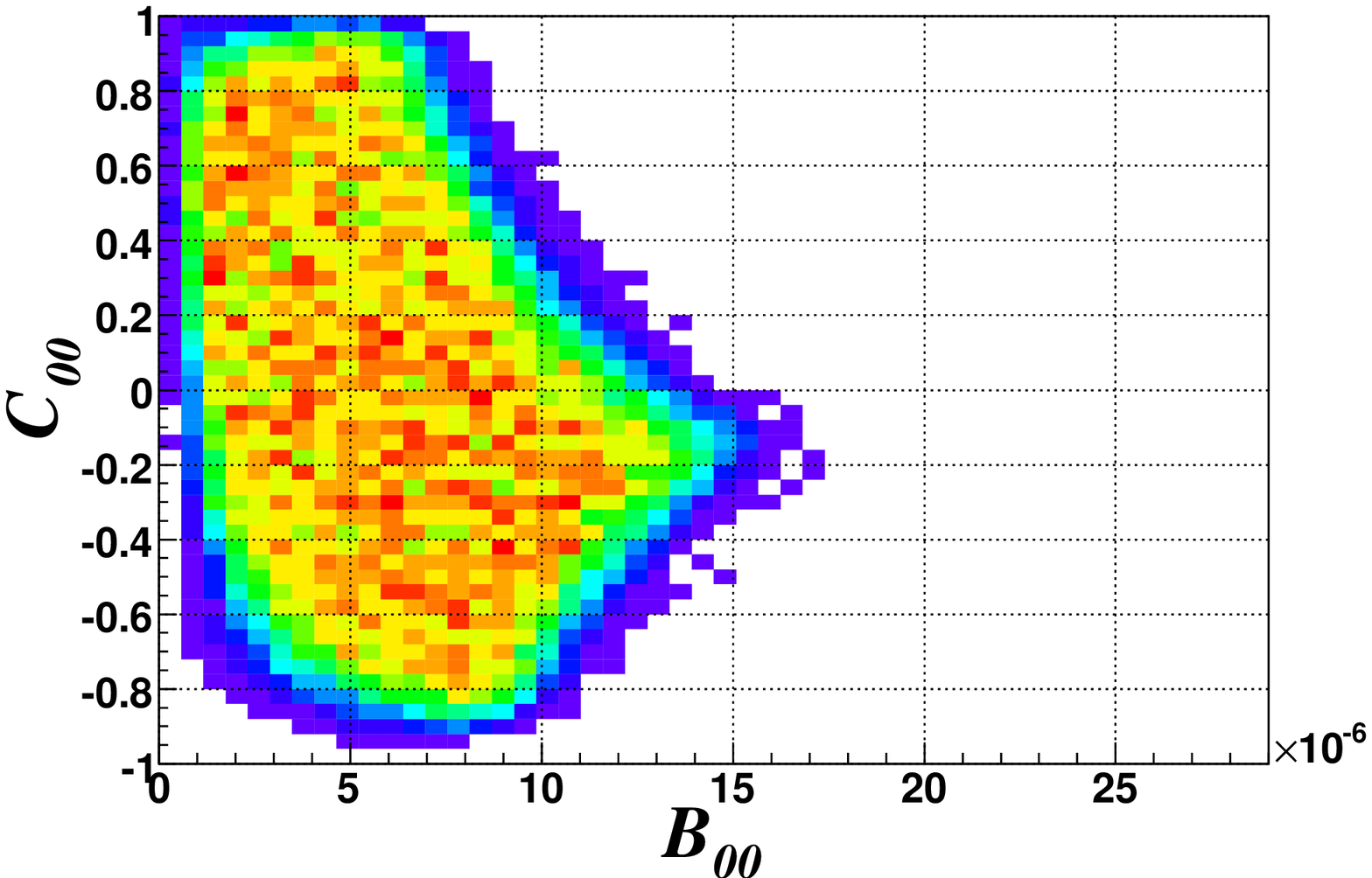}
  \caption{\label{fig:nopi0pi0} Left: Posterior distribution for the '1i' parameterization when \Coo and \Boo are 
uniformly sampled in [-1, +1] and [0, 20\Boo], respectively. Sampling is identical to figure \ref{fig:posterior}. Right: Joint distribution of \Boo\ and \Coo\ implied by the observations and the assumption of isospin symmetry.}
}

\section{Conclusions}

In the debate concerning the analysis of the CKM phase \al\ problem we
have contributed two important points.  The first is a formulation of
the problem that is invariant to the choice of parameterization.  The
second is the correct interpretation of the posterior marginal
distribution for \al\ as a representation of our state of knowledge.

In the CKM Phase $\alpha$ problem the relationships between the
parameters of the model and the observables is deterministic.  In this
case the appropriate statistical technique to find the distribution
over the model parameters is that of the
transformation-of-variables.  This gives us a distribution over the
model parameters that summarizes our state of knowledge.  It does not,
and cannot, tell us if our model is true or false.  We have no way of
knowing the actual mechanisms of the external universe.  We can only
generate models of the universe and use data to cast light on these
models.  However ``true'' we may think our models are today, better
models will certainly be developed tomorrow.  The scientific method is
composed of the cycle of model formulation, testing against
observations, and model revision and development.  Bayesian statistics
provides many tools to facilitate this process.

\begin{acknowledgments}
RDM is supported by the NASA AISR program.
The authors thank St\'ephane T'Jampens, Roger Barlow and Louis Lyons
for helpful discussions. 
\end{acknowledgments}

\appendix

\section{\label{app:invariance}Reparameterization invariance of the marginal posterior pdf
  over \al\ }

We consider a system of $N$ random variables $X_i$ ($i=1...N$), which are
related to a set of $N$  
observables ${\cal O}_i$ as ${{\cal O}_i}=f_i(\mathbf{X})$. 
We also assume that it is possible to reparameterize the variables
$X_i$ into a set $Y_i$ so that  
$X_1=Y_1=\alpha$, $Y_i=\phi_i(\mathbf{X})$, and ${\cal O}_i=g_i(\mathbf{Y})$.
Within the Bayesian framework, we consider a dataset ${d}$ used to
estimate the observables, which yields 
the posterior pdf $p_{\mathbf{\cal O}}(\mathbf{o}|d)$. Under the
further hypothesis that $f$, $g$ and $\phi$ are 
invertible, we can write the marginal posterior on $\alpha$ using the
parameterization $Y$ as: 
\begin{eqnarray}
 p^{\mathbf{Y}}_\alpha(\alpha|{d}) &=& \int...\int p_{\mathbf{\cal
     O}}(\mathbf{o}|d) |J_g| dy_2...dy_N\, \mbox{as}\,
     p_\mathbf{X}(\mathbf{x}|{d})=p_{\mathbf{\cal O}}(\mathbf{o}|d)
     |J_g|\\ 
&=& \int...\int p_{\mathbf{\cal O}}(\mathbf{o}|d) |J_g| |J_\phi| dx_2...dx_N\\
&=& \int...\int p_{\mathbf{\cal O}}(\mathbf{o}|d) |J_f| dx_2...dx_N\\
&=& p^{\mathbf{X}}_\alpha(\alpha|{d}) \, \mbox{as }\, p_\mathbf{Y}(\mathbf{y}|{d})=p_{\mathbf{\cal O}}(\mathbf{o}|d) |J_f|,
\end{eqnarray}
proving that the marginal posterior on $\alpha$ is parameterization
invariant. Thus, if a Bayesian analysis has been performed on the
dataset ${d}$ so that the posterior pdf on the observables is known,
the marginal posterior on $\alpha$ obtained by the change of variables
$Y_i=\phi_i(\mathbf{X})$
is invariant under reparameterization
of the $N-1$ marginalized variables $X_i$, $i=2\ldots N$.

\section{\label{app:parameterizations}Parameterizing the CKM Phase \al\ Problem}
We give details here of the three parameterizations, the
Pivk-LeDiberder (PLD), the Explicit Solution (ES) and the 1i
parameterizations.

\subsection{The Pivk-LeDiberder Parameterization}

PLD introduces six parameters, $\alpha, \alpha_{eff}, \mu, a,
\bar{a},\Delta$, via 
\begin{eqnarray}
A^{+-} = \mu a \  &,& \  \bar{A}^{+-} =  \mu \bar{a} e^{2i\alpha_{eff}}  \nonumber\\
A^{+0}  =  \mu e^{i(\Delta-\alpha)} &,&  \bar{A}^{+0}  =  \mu  e^{i(\Delta+\alpha)} \\
A^{00}  = \mu e^{i(\Delta-\alpha)}\left( 1-\frac{a}{\sqrt{2}} e^{-i(\Delta-\alpha)}\right) &,&  
\bar{A}^{00} =  \mu e^{i(\Delta+\alpha)}\left(1-\frac{\bar{a}}{\sqrt{2}}e^{-i(\Delta+\alpha-2\alpha_{eff})}\right)
\nonumber 
\end{eqnarray}
which results in
\begin{eqnarray}
\Bpm & = & C  \frac{\tau_{B^0}}{2}  \mu^2(a^2+\bar{a}^2)\nonumber \\
\Boo & = & C  \frac{\tau_{B^0}}{2}  \mu^2\left(2+\frac{a^2+\bar{a}^2}{2}-\sqrt{2}(a\cos{(\Delta-\alpha)}+\bar{a}\cos{(\Delta+\alpha-2\alpha_{eff})})\right)\nonumber \\
\Bpo & = & C  \tau_{B^+}  \mu^2\nonumber \\
\Cpm & = & \frac{ a^2-\bar{a}^2 }{ a^2+\bar{a}^2 }\label{eq:pld} \\
\Coo & = & \frac{\frac{a^2-\bar{a}^2}{2}-\sqrt{2}(a\cos{(\Delta-\alpha)}-\bar{a}\cos{(\Delta+\alpha-2\alpha_{eff})})}{2+\frac{a^2+\bar{a}^2}{2}-\sqrt{2}(a\cos{(\Delta-\alpha)}+\bar{a}\cos{(\Delta+\alpha-2\alpha_{eff})})}\nonumber \\
\Spm & = & 2\frac{a\bar{a}}{a^2+\bar{a}^2}\sin{2\alpha_{eff}} \nonumber
\end{eqnarray}
where $C=(16\pi M_B\hbar)^{-1}$.
This system can be solved to give
\begin{eqnarray}
 \mu^2 &=&\frac{\Bpo}{C\tau_{B^+}}\nonumber \\
a^2 &=& K(1+\Cpm)\nonumber \\
\bar{a}^2 &=& K(1-\Cpm)\nonumber \\
\sin{2\alpha_{eff}} &=& \frac{\Spm}{\sqrt{1-(\Cpm)^2}} \equiv
\sin{s} \label{eq:pldinv} \\
 \cos{(\Delta-\alpha)} &=&
 \frac{(1+\Cpm)K-2K\frac{\Boo}{\Bpm}(1+\Coo)+2}{2\sqrt{2K(1+\Cpm)}} \equiv \cos{t}\nonumber \\ 
 \cos{(\Delta+\alpha-2\alpha_{eff})} &=&
 \frac{(1-\Cpm)K-2K\frac{\Boo}{\Bpm}(1-\Coo)+2}{2\sqrt{2K(1-\Cpm)}} \equiv \cos{u}\nonumber  
\end{eqnarray}
where we define
$\displaystyle{K=\frac{\Bpm}{\Bpo}\frac{\tau_{B^+}}{\tau_{B^0}}}$,
and $s,t$ and $u$ as in the final three equations.
The fourth equation  yields $2\alpha_{eff}=s$ or
$2\alpha_{eff}=\pi - s$. The final two equations  yield
$\Delta+\alpha=\epsilon t + s$ or $\Delta+\alpha=\epsilon t + \pi -s$
and 
$\Delta-\alpha=\epsilon' u$ or $\Delta+\alpha=\epsilon t + \pi-s$,
respectively, where $\epsilon,\epsilon'=\pm1$. Finally, we obtain
$\alpha=\epsilon t +\epsilon' u + s$ or $\alpha=\epsilon t +\epsilon'
u + \pi - s$ as the 8 solutions corresponding to each set of values of
the observables.

\subsection{The Explicit Solution Parameterization}

The Explicit Solution (ES) parameterization \cite{pivk2004} begins
with the same parameters as the PLD parameterization, and then defines
\begin{eqnarray*}
  c & = & \cos(\phi),\, \phi = \al - \Delta \\
  \bar{c} & = & \cos(\bar{\phi}),\, \bar{\phi} =
  \al+\Delta-2\alpha_{eff}
\end{eqnarray*}
and also $s=\sin(\al)$, $\bar{s}=\sin(\bar{\phi})$.
Using the identity $2\al = 2\alpha_{eff} + \phi +\bar{\phi}$ allows
the following solution to be derived.
\begin{eqnarray}
\tan \alpha & = &
\frac{\sin(2\alpha_{eff})\bar{c}+\cos(2\alpha_{eff})\bar{s}+s}
{\cos(2\alpha_{eff})\bar{c}-\sin(2\alpha_{eff})\bar{s}+c} \nonumber \\
\sin(2\alpha_{eff}) & = & \frac{\Spm}{\sqrt{1-\Cpm{}^2}} \nonumber \\
\cos(2\alpha_{eff}) & = & \pm \sqrt{1-\sin^2(2\alpha_{eff})} \nonumber \\
c & = & \sqrt{\frac{\tau_{B^+}}{\tau_{B^0}}} \frac{
    \frac{\tau_{B^0}}{\tau_{B^+}} \Bpo+\Bpm(1+\Cpm)/2 - \Boo(1+\Coo)}
  {\sqrt{2\Bpm\Bpo(1+\Cpm)}}
\nonumber \\
\bar{c} & = & \sqrt{\frac{\tau_{B^+}}{\tau_{B^0}}} \frac{
    \frac{\tau_{B^0}}{\tau_{B^+}} \Bpo+\Bpm(1-\Cpm)/2 - \Boo(1-\Coo)}
  {\sqrt{2\Bpm\Bpo(1-\Cpm)}}
\nonumber \\
s & = & \pm\sqrt{1-c^2} \nonumber \\
\bar{s} & = & \pm\sqrt{1-\bar{c}^2} 
\label{eq:es}
\end{eqnarray}
where the 8 solutions in the range $[0,\pi ]$ are apparent from the
three arbitrary signs.

\subsection{The 1i Parameterization}

Botella and Nebot introduce the following parameterization
\begin{eqnarray*}
  A^{+-}          & = & e^{-i\al}\Ttt (T+iP) \\
  \sqrt{2} A^{+0} & = & e^{-i\al}\Ttt \\
  \bar{A}^{+-}    & = & e^{+i\al}\Ttt (T-iP) \\
  \sqrt{2} A^{00} & = & e^{-i\al}\Ttt (1-T-iP) \\
  \sqrt{2} \bar{A}^{+0} & = & e^{+i\al}\Ttt \\
  \sqrt{2} \bar{A}^{00} & = & e^{+i\al}\Ttt (1-T-iP)
\end{eqnarray*}
and writing $T$ and $P$ in terms of real and imaginary parts allows
the system of equations for the observables to be inverted in terms of
\al, \Ttt, $T_r$, $T_i$, $P_r$, $P_i$, in the following way :
\begin{eqnarray*}
T&=&\sqrt{\frac{2\Bpo}{\taup C}}\\
T_r &=& \frac{2\,\Bpo\,\tauz+\left( \Bpm-2\,\Boo\right) \,\taup}{4\,\Bpo\,\tauz}\\
P_i &=& \frac{\left( 2\,\Boo\,\Coo-\Bpm\,\Cpm\right) \,\taup}{4\,\Bpo\,\tauz}\\
(T_i+P_r)^2 &=& \frac{\Bpm \taup}{2\Bpo\tauz(1+\Cpm)} - (T_r-P_i)^2\\ 
(T_i-P_r)^2&=& \frac{\Bpm \taup}{2\Bpo\tauz(1-\Cpm)} - (T_r+P_i)^2\\
\alpha&=&\arctan\left(\pm\frac{\sqrt{b^2+a^2-c^2}+ a}{c-b}\right)
\end{eqnarray*}
\noindent with $a=(T_i^2-P_i^2+T_r^2-P_r^2),\,b=2P_iT_i+2P_rT_r,\,c=\Spm\Bpm\taup/(2\tauz\Bpo)$.

\end{document}